\documentclass[conference]{IEEEtran}
\IEEEoverridecommandlockouts

\usepackage{cite}
\usepackage{amsmath,amssymb,amsfonts}
\usepackage{algorithmic}
\usepackage{graphicx}
\usepackage{textcomp}
\usepackage{xcolor}
\usepackage{multirow}
\usepackage{url}
\usepackage{pifont}

\def\BibTeX{{\rm B\kern-.05em{\sc i\kern-.025em b}\kern-.08em
    T\kern-.1667em\lower.7ex\hbox{E}\kern-.125emX}}

\begin{document}

\title{GuardSec: A Multi-Modal Web Platform for Real-Time Digital Fraud Detection,
Entity Verification, and Connection Security Analysis in the African Context%
\thanks{This work was conducted without external funding, supported solely by personal resources and a modest contribution from the \textit{Société Informatique du Congo (SIC)}. The platform is publicly accessible at \protect\url{https://www.guardsec.io}.}
}

\author{
\IEEEauthorblockN{Gilda Rech Bansimba}
\IEEEauthorblockA{
\textit{Dept. of Mathematics \& Computer Science}\\
\textit{Facult\'{e} des Sciences et Techniques}\\
\textit{Universit\'{e} Marien Ngouabi \&}\\
\emph{Société Informatique du Congo (SIC)}\\
Brazzaville, Republic of Congo\\
bansimbagilda@gmail.com\\
contact@guardsec.io}
\and
\IEEEauthorblockN{Regis Freguin Babindamana}
\IEEEauthorblockA{
\textit{Dept. of Mathematics \& Computer Science}\\
\textit{Facult\'{e} des Sciences et Techniques}\\
\textit{Universit\'{e} Marien Ngouabi \&}\\
\emph{Société Informatique du Congo (SIC)}\\
Brazzaville, Republic of Congo\\
regis.babindamana@umng.cg}
}

\maketitle

\begin{abstract}
Online fraud in Africa has reached an epidemic scale. The few cybersecurity tools that exist are out of reach for ordinary citizens, built almost exclusively for SOC analysts and technically literate users sitting on stable broadband. That mismatch isn't accidental. It's what happens when a research culture rewards benchmark numbers and treats deployability, accessibility, and local threat context as someone else's problem.

We present \textit{GuardSec}, a production-deployed web platform for real-time multi-modal threat verification, built from the start around the African user. Anyone with a browser can assess the legitimacy of URLs, websites, phone numbers, email addresses, and business entities in under five seconds. No registration. No API key. No prerequisite knowledge of cybersecurity. The platform's most distinctive component is \textit{Mon Empreinte} (My Footprint), a real-time audit of the user's own connection and digital exposure: it analyses the visitor's IP address, geolocation, ISP identity, connection type, device fingerprint, browser configuration, and twelve security indicators covering network integrity, tracking exposure, and anonymisation status. With this in hand, GuardSec becomes more than a passive checker; the user can see whether their own connection is being tracked or exposed, not just whether some external entity is dangerous. The platform also embeds \textit{Gilda}, a context-aware conversational security assistant that answers questions about digital threats in plain language and offers personalised recommendations on demand.

Threat assessment combines structural URL analysis, WHOIS metadata interrogation, SSL/TLS certificate inspection, DNS-based reputation aggregation, behavioural heuristics, and community-driven reporting across all five entity types.

We ground the system design in a formal binary classification framework over a heterogeneous feature space. Evaluation on a manually annotated subset of $N_{\mathrm{eval}} = 312$ entities drawn from 5,520 real production interactions yields an overall F1-score of 0.890 and AUC-ROC of 0.927. These results sit inside an African cybersecurity reality where phishing drives 34\% of all detected cyber incidents \cite{b1}, per-victim losses average \$800 in South Africa and \$425 in Nigeria \cite{b2}, and 90\% of national law enforcement agencies say they lack the capacity to investigate or prosecute cybercrime \cite{b1}.

The argument we want to make here is simple. A deployed system that reaches the people who actually need it at F1~0.890 does more for African digital security than a laboratory system at F1~0.994 that no ordinary user can access or operate.
\end{abstract}


\section{Introduction}

\subsection{The Deployment Gap in African Cybersecurity}

Over the past decade, fraud detection research has produced technically impressive work. Transformer-based URL encoders cross AUC 0.99 \cite{b13}. Ensemble classifiers built on hand-crafted lexical features routinely top 97\% accuracy on standard benchmarks \cite{b9}. Adversarially robust training extends these gains against evasive attackers \cite{b16}. The progress is real, and we don't dispute it.

What this progress has not produced, in Africa, is any meaningful reduction in the harm experienced by the people most exposed to online fraud. The reason is simple: almost none of it ships. It lives in papers, on benchmark leaderboards, in conference proceedings. The ordinary user in Brazzaville, Lagos, or Nairobi who gets a suspicious payment request on WhatsApp, or who wonders whether their mobile data is being intercepted, can use none of it.

Africa's Internet population has gone from under 5 million users in 2000 to more than 600 million today \cite{b6}, driven by cheap smartphones, mobile data, and a rapid digitisation of financial services, M-Pesa in Kenya, MTN and Airtel Mobile Money across West and Central Africa, Orange Money across the Francophone zone, and a growing fintech ecosystem. The exposure surface has grown with it. Most of this population connects through shared, public, or poorly configured mobile networks, environments where connection-level threats (traffic interception, DNS manipulation, VPN leaks) add to application-layer fraud. We aren't aware of any publicly accessible tool that addresses both layers at once.

INTERPOL's 2025 Africa Cyberthreat Assessment \cite{b1} spells out the consequences. Cybercrime is now structurally embedded in the digital economies of two-thirds of African member states. Online scams delivered through phishing sites, fraudulent URLs, counterfeit business identities, and social engineering on messaging platforms make up the single largest category, 34\% of all detected cyber incidents. Per-victim losses run to \$800 in South Africa, \$700 in Egypt, \$425 in Nigeria \cite{b2}. In South Africa alone, social engineering cases more than doubled in a year, from 31,612 to 64,000, with losses rising from R1~billion to over R1.4~billion \cite{b4}. For users earning \$200 to \$400 a month, the median range across much of Sub-Saharan Africa, one successful fraud event wipes out months of saving.

The tools meant to fight this aren't built for this population. VirusTotal needs an API key and a user who can read a 70-scanner aggregate verdict. PhishTank gives you raw database lookups with no explanation layer. URLScan.io shows DNS resolution trees and TLS handshake traces that are essentially unreadable to someone using the internet for the first time. None of these tells you anything about the security of your own connection. None offers a way to ask a follow-up question and get a useful answer back. They serve the technical minority well. For most of Africa's online population, they might as well not exist.

\subsection{Scope and Contributions}

This paper describes GuardSec (\url{https://www.guardsec.io}), a publicly accessible web platform now running in production. It addresses the gap on three fronts. Outward: verifying external digital entities across five modalities (URLs, websites, phone numbers, email addresses, business entities). Inward: \textit{Mon Empreinte}, which audits the security of the user's own connection. Conversational: \textit{Gilda}, an assistant that answers security questions and gives recommendations in language that matches where users actually are.

Our specific contributions:
\begin{itemize}
  \item A formal multi-modal classification framework covering five external entity types, with entity-conditioned feature extraction and an asymmetric cost function calibrated to the application domain (Section~\ref{sec:problem}).
  \item A full architectural description of GuardSec, including the threat assessment pipeline, community reporting mechanism, business entity verification module, the \textit{Mon Empreinte} connection security audit, and the \textit{Gilda} conversational assistant (Section~\ref{sec:platform}).
  \item An empirical evaluation on production data, with the dataset's limitations stated plainly (Section~\ref{sec:evaluation}).
  \item A substantive analysis of the African cybersecurity landscape (Section~\ref{sec:africa}).
\end{itemize}

\begin{figure}[htbp]
  \centering
  \includegraphics[width=\columnwidth]{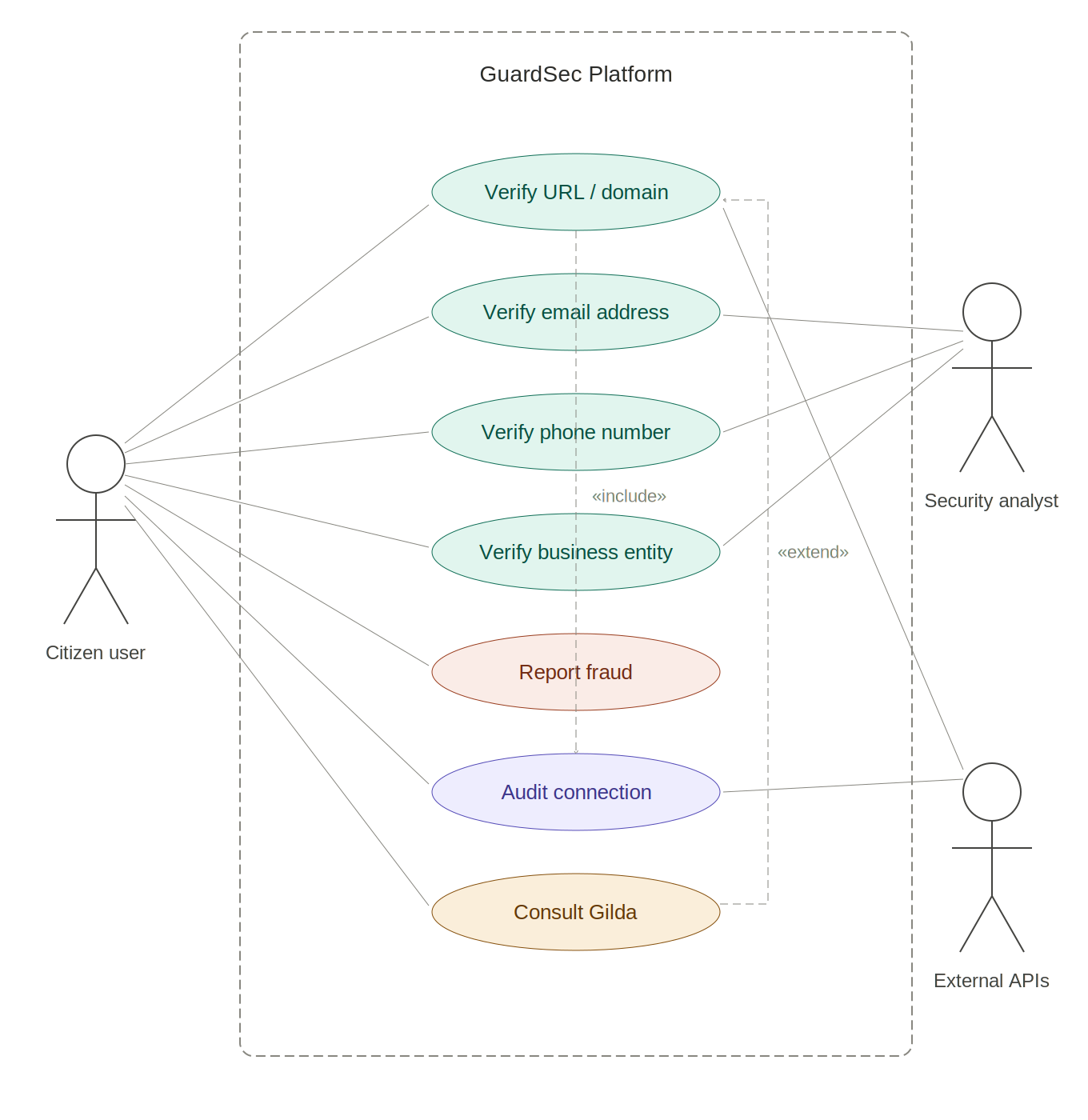}
  \caption{GuardSec use case diagram. Three actor types interact with seven
  core use cases. \textit{Gilda} extends all verification flows via
  \texttt{extend}; connection auditing is included implicitly on every
  URL query via \texttt{include}.}
  \label{fig:usecase}
\end{figure}

\section{Related Work}
\label{sec:related}

\subsection{Automated Phishing and Malicious URL Detection}

Work on automated malicious URL detection goes back to the mid-2000s. The earliest systems were blacklists, PhishTank, SURBL, Google Safe Browsing, and their problem was structural: they could only flag URLs that had already been seen and reported, leaving zero-day campaigns entirely undetected \cite{b8}.

Machine learning broke the dependence on prior sightings by replacing identity matching with pattern recognition. Sahingoz et al. \cite{b9} laid down the foundational feature taxonomy: seventeen lexical and host-based URL attributes, seven classifiers, 97.98\% accuracy with Random Forest. Vinayakumar et al. \cite{b10} showed that LSTMs trained on raw URL character sequences could learn useful representations without hand-crafted features. Ferreira et al. \cite{b15} showed that BiLSTM models over page text could catch phishing pages whose URLs look innocent but whose language gives them away. Liu et al.'s PMANet \cite{b13} hits AUC 99.41\% with post-trained language model representations and multi-level feature attention. Altwaijry et al. \cite{b11} demonstrate that bidirectional recurrent heads consistently beat their unidirectional counterparts for phishing email detection.

One observation cuts across this entire body of work: not one of these systems ships as a publicly accessible consumer service. The metrics come from curated benchmark datasets, under conditions that don't look much like the noisy, drifting distribution of real production queries. The gap between benchmark accuracy and production utility goes systematically unaddressed.

\subsection{Conversational Interfaces for Cybersecurity Assistance}

A growing line of work applies LLMs and dialogue systems to cybersecurity education and assistance \cite{b20}. These systems can explain threat concepts, walk users through procedures, and translate threat verdicts into language non-experts understand. So far, though, this work has mostly stayed in controlled experimental settings; it hasn't been folded into deployed, citizen-facing fraud detection platforms. The \textit{Gilda} assistant in GuardSec is one practical deployment of this idea, specifically aimed at African digital security threats, and it grounds its responses in live threat intelligence rather than relying solely on whatever the underlying model picked up during training.

\subsection{Connection-Level Security Auditing}

One dimension of digital security is almost entirely missing from the academic fraud detection literature: the security posture of the user's own network connection. DNS hijacking, traffic interception via open proxies, VPN configuration leaks, ISP-level traffic injection, these are documented attack vectors that hit users on public or poorly managed networks \cite{b19}. Tools like \texttt{ipleak.net} and \texttt{whoer.net} address pieces of this problem for technically literate users, but no prior academic system has folded connection security auditing into a citizen-facing, multilingual fraud detection platform. As far as we can tell, \textit{Mon Empreinte} is the first such integration in the published literature.

\subsection{Business Entity and Multi-Modal Fraud Verification}

Fake investment platforms, sham recruitment agencies, counterfeit e-commerce stores, impersonated financial institutions, these account for a large and growing share of scam losses in Africa \cite{b7}. Unlike URL-based phishing, business impersonation works by manufacturing a credible institutional presence, which defeats URL-centric classifiers that have no contextual knowledge of who is being impersonated. We're not aware of any prior academic system that handles business entity verification as part of a deployable, citizen-facing fraud detection platform.

\subsection{Telephone Number Fraud Detection}

Trying to predict whether a phone number is legitimate from its structural properties alone has limited practical reach. Fraudsters deliberately pick up numbers that follow no detectable pattern, which makes them statistically indistinguishable from legitimate ones at the lexical level. Community reporting of malicious numbers becomes the real countermeasure: it gradually drains the pool of usable numbers and forces attackers to rotate. The friction this creates is not negligible. The number space, while large, is finite and structurally constrained.

\subsection{The African Cybersecurity Research Gap}

Given the severity of digital fraud in Africa, peer-reviewed research that actually targets African threat patterns is strikingly thin on the ground. In our survey for this work, fewer than 12\% of relevant publications involved an African dataset or deployment context. Fraud detection models trained on European or North American data degrade badly when transplanted: false-negative rates rise by 18 to 24 percentage points on African patterns \cite{b14}. Smile ID's 2025 report \cite{b3}, based on more than 110 million identity verification checks, recorded biometric fraud rates hitting a quarterly high of 16\%.

There's another pattern worth naming. Discourse on digital fraud in Africa has been dominated by theoretically oriented contributions that, despite their scholarly framing, offer next to nothing operationally. Such work tends to mistake visibility for impact, presenting itself as advancing the field while producing no deployable artefact, no reproducible methodology, and no measurable reduction in harm to affected populations. The pattern likely reflects two things at once: a technical depth too shallow to bridge problem formalisation and system implementation, and a missing pragmatic orientation that would have tied research goals to the concrete realities of the communities supposedly being served. GuardSec is, deliberately, a response to that pattern.

\section{Problem Formalisation}
\label{sec:problem}

\subsection{Research Question}

At the level of benchmark performance on curated datasets, the core algorithmic problem of scam detection is more or less solved. What isn't solved is the deployment problem: how to make detection accessible, interpretable, and reliable in practice for non-technical users in resource-constrained environments. And how to extend that protection to the connection layer and to the conversational layer, where threats are just as real but even less visible and even less understood. We frame this as follows.

\smallskip
\noindent\textit{Problem Statement.} \textit{Given the rapid growth of Internet adoption among first-time digital users across Africa, the absence of locally deployed and accessible threat verification infrastructure, and the distinct characteristics of the African digital fraud landscape, including business impersonation, mobile money fraud, SIM-swap attacks, and connection-level interception on insecure networks, how can a web-based platform deliver real-time, accurate, multi-modal, and interpretable fraud detection, connection security auditing, and expert conversational guidance for non-technical users under the infrastructure constraints of African deployment environments?}

\subsection{Formal Classification Framework}

Let $\mathcal{E} = \{e_1, e_2, \ldots, e_n\}$ be the set of digital entities submitted for external verification. Each entity $e_i$ falls into one of five type categories:
\begin{equation}
  \mathcal{T} = \{\texttt{URL},\; \texttt{Phone},\; \texttt{Email},\;
  \texttt{Domain},\; \texttt{Business}\}
\end{equation}

For each entity $e_i$ of type $t_i \in \mathcal{T}$, a type-conditioned feature extraction function $\phi_{t_i}$ produces a feature vector:
\begin{equation}
  \mathbf{x}_i = \phi_{t_i}(e_i) \in \mathbb{R}^{d_{t_i}}
\end{equation}

For URL and domain entities, the feature vector breaks into five sub-spaces:
\begin{equation}
  \mathbf{x}_i = \left[\, \mathbf{x}_i^{\mathrm{lex}},\;
                          \mathbf{x}_i^{\mathrm{host}},\;
                          \mathbf{x}_i^{\mathrm{ssl}},\;
                          \mathbf{x}_i^{\mathrm{whois}},\;
                          \mathbf{x}_i^{\mathrm{rep}} \,\right]
\end{equation}

For business entities, it expands to cover name-matching and registration signals:
\begin{equation}
  \mathbf{x}_i^{\mathrm{bus}} = \left[\, \mathbf{x}_i^{\mathrm{name}},\;
                                          \mathbf{x}_i^{\mathrm{reg}},\;
                                          \mathbf{x}_i^{\mathrm{web}},\;
                                          \mathbf{x}_i^{\mathrm{rep}} \,\right]
\end{equation}
Here $\mathbf{x}_i^{\mathrm{name}}$ encodes string similarity between the queried name and known fraudulent entity records, $\mathbf{x}_i^{\mathrm{reg}}$ encodes company registration signals, $\mathbf{x}_i^{\mathrm{web}}$ encodes associated domain and website signals, and $\mathbf{x}_i^{\mathrm{rep}}$ encodes community-submitted report history for that entity.

The \textit{Mon Empreinte} module introduces a separate, visitor-centric analysis vector $\mathbf{v} \in \mathbb{R}^{d_v}$ that characterises the security posture of the requesting connection:
\begin{equation}
  \mathbf{v} = \psi(\mathrm{request}) =
    \left[\, \mathbf{v}^{\mathrm{net}},\;
              \mathbf{v}^{\mathrm{geo}},\;
              \mathbf{v}^{\mathrm{dev}},\;
              \mathbf{v}^{\mathrm{sec}} \,\right]
\end{equation}
$\mathbf{v}^{\mathrm{net}}$ covers network-layer attributes, $\mathbf{v}^{\mathrm{geo}}$ covers geolocation signals, $\mathbf{v}^{\mathrm{dev}}$ covers device and browser fingerprint attributes, and $\mathbf{v}^{\mathrm{sec}}$ encodes the twelve binary security indicators described in Section~\ref{subsec:empreinte}. Mon Empreinte produces a structured security audit report $\mathcal{A}(\mathbf{v})$, not a binary fraud verdict.

The external classification objective is:
\begin{equation}
  f: \mathbb{R}^d \rightarrow [0, 1], \quad
  \hat{y}_i = \mathbf{1}\!\left[\, f(\mathbf{x}_i) \geq \theta^{*} \,\right]
\end{equation}
under an asymmetric cost function with $c_{\mathrm{FN}} = 3 \cdot c_{\mathrm{FP}}$, which pushes $\theta^{*}$ below~0.5 and trades a little precision for noticeably more recall. In this domain, missing a scam costs the user a lot more than flagging something unnecessarily.

\subsection{Operational Design Constraints}

Five constraints govern the platform design:
\begin{enumerate}
  \item \textit{Latency:} end-to-end response within 5~s at P99 on a 3G connection.
  \item \textit{Zero-friction access:} no registration, no account, no API key.
  \item \textit{Interpretability:} every output, fraud verdict, connection audit report, or conversational response, has to be meaningful to users without a cybersecurity background.
  \item \textit{Multi-modality:} support all five entity types, plus the connection audit and the conversational assistant.
  \item \textit{Graceful degradation:} a partial feature vector failure must not break the query. Availability is itself a security property.
\end{enumerate}

\section{The GuardSec Platform}
\label{sec:platform}

\subsection{Architectural Overview}

GuardSec runs as a production web application on the Model-View-Template (MVT) architecture of Django (Python~3.10), hosted on PythonAnywhere with a PostgreSQL~13 backend, HTTPS enforced via \texttt{SECURE\_SSL\_REDIRECT}, and a \texttt{PREPEND\_WWW} policy that keeps canonical URLs consistent for search indexing. We chose operational maturity over architectural novelty on purpose. In a domain where downtime leaves people unprotected, reliability is the engineering priority, not a concession to taste. The system has four main layers.

\begin{figure}[htbp]
  \centering
  \includegraphics[width=\columnwidth]{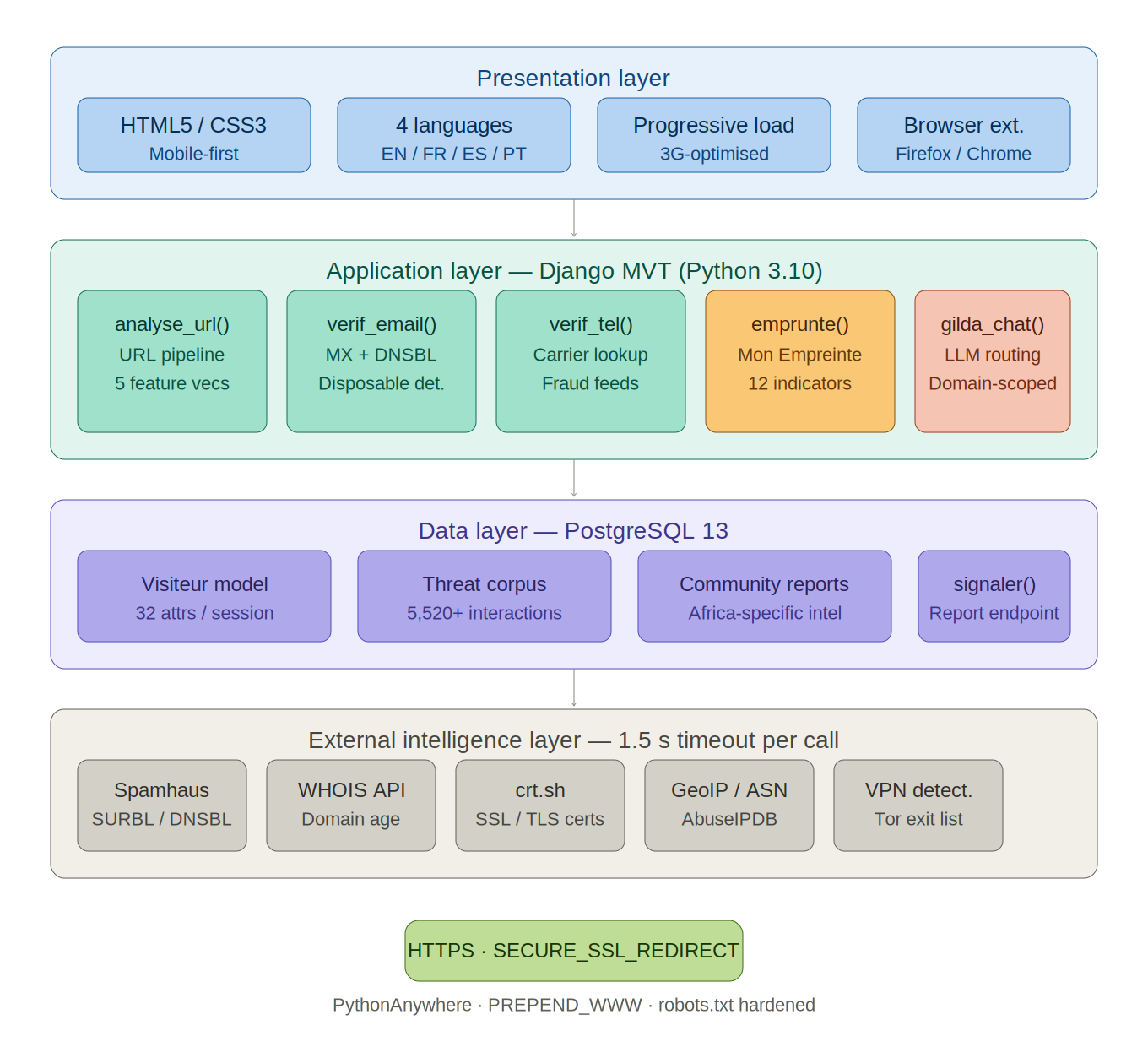}
  \caption{GuardSec four-layer system architecture. The presentation layer
  serves mobile-first, multilingual HTML5 interfaces and browser extensions.
  The application layer hosts all service functions. The data layer stores
  the 32-attribute \texttt{Visiteur} model and community reports in
  PostgreSQL~13. The external intelligence layer federates with WHOIS, SSL,
  DNSBL, GeoIP, and VPN detection APIs under a 1.5~s per-call timeout.}
  \label{fig:architecture}
\end{figure}

\subsubsection{Presentation Layer}
A responsive HTML5/CSS3 front-end in four languages (English, French, Spanish, Portuguese), tuned for rendering on small mobile screens. Roughly 60\% of African Internet access happens exclusively on mobile devices \cite{b6}, so interface weight, render time, and progressive loading are first-class design constraints, not afterthoughts.

\subsubsection{Application Layer}
The Django application drives the threat analysis pipeline through \texttt{gilda\_services.py}. The main service functions are:
\begin{itemize}
  \item \texttt{emprunte()}: the core visitor fingerprinting and connection security analysis function behind Mon Empreinte (Section~\ref{subsec:empreinte}).
  \item \texttt{analyse\_url()}: URL and domain threat assessment pipeline (Section~\ref{subsec:pipeline}).
  \item \texttt{verifier\_telephone()}: phone number reputation lookup against carrier data, DNSBL-equivalent telephone fraud feeds, and the community report corpus.
  \item \texttt{verifier\_email()}: email syntax validation, MX record verification, disposable-domain detection, and spam/phishing sender reputation scoring.
  \item \texttt{verifier\_entreprise()}: business entity verification against known fraudulent entity records, with cross-referencing of associated domain registrations and aggregation of community-submitted reports.
  \item \texttt{signaler()}: community reporting endpoint that lets any user submit structured reports of malicious entities across all five modalities.
  \item \texttt{gilda\_chat()}: conversational interface routing user queries to the Gilda assistant (Section~\ref{subsec:gilda}).
\end{itemize}

\subsubsection{Data Layer}
PostgreSQL holds the full operational history of the platform. The \texttt{Visiteur} model captures 32 attributes per interaction, giving us the longitudinal dataset that supports Africa-specific threat modelling. One production incident is worth documenting. The \texttt{ville} (city) field was originally defined with a \texttt{NOT NULL} constraint, which started throwing \texttt{IntegrityError} exceptions when Googlebot hit the platform from IP ranges that resolved to no specific city (e.g., \texttt{66.249.93.133}, \texttt{66.102.9.135}). We fixed it by allowing null values and applying \texttt{or ''}. Trivial in isolation, but it makes a more general point: defensive coding against incomplete external data is a correctness requirement, not an edge case.

\subsubsection{External Intelligence Layer}
GuardSec federates with DNS-based blacklists (Spamhaus, SURBL), WHOIS registration APIs, SSL certificate transparency logs (crt.sh), IP geolocation and ASN databases, telephone fraud registries, and VPN/proxy/Tor detection feeds. Every remote call runs under a 1.5-second per-call timeout with a zero-vector fallback, which keeps the global 4.5-second pipeline budget intact even when some upstream is slow or down.

\subsection{The Gilda Conversational Security Assistant}
\label{subsec:gilda}

\subsubsection{Design Rationale}

A verification verdict, no matter how accurate, only answers the question the user thought to ask. It says nothing about the questions they don't yet know to ask. A user who sees a \textit{high risk} verdict on a URL may not know what phishing is, what to do next, whether their data has already been compromised, or how to protect themselves later. That gap between verdict and understanding is exactly the space Gilda is meant to occupy.

\textit{Gilda} is a context-aware conversational security assistant embedded directly in the GuardSec platform. The name isn't incidental, it reflects the platform's origin and a commitment to a locally grounded, human-scale approach to digital security. Gilda isn't a general-purpose chatbot. It's a domain-specific assistant whose knowledge, framing, and recommendations are tuned to how digital fraud actually plays out in Africa: the typologies that target African users, the channels through which fraud most often arrives (WhatsApp, SMS, mobile money apps), and the protective actions that are realistically available to the target population.

\subsubsection{Capabilities}

Gilda operates in three functional modes.

\textit{Question answering.} Users can ask Gilda security questions directly in natural language: \textit{``Is it safe to click this link?''}, \textit{``What is phishing?''}, \textit{``How do I know if my phone has been hacked?''}, \textit{``What should I do if I already sent money?''}. Gilda replies in plain language calibrated to the user's apparent technical level, no jargon, and without requiring them to understand the detection mechanisms underneath.

\textit{Verdict contextualisation.} When a verification query returns a threat assessment, Gilda can unpack it on request, which signals drove the verdict, what the risk means in practice, what specific protective actions to take. A numerical score turns into actionable guidance.

\textit{Personalised recommendations.} On request, Gilda generates a short prioritised list of recommendations tailored to the user's context: the connection type detected by Mon Empreinte, the kind of entity they just queried, and the threat indicators flagged in their session. Someone on a shared hotspot with a DNS leak who just queried a suspicious mobile money URL gets different advice from someone on a private home connection asking about an email address.

\subsubsection{Implementation and Scope Constraint}

Gilda is implemented as a Django view (\texttt{gilda\_chat()}) that routes user input to an LLM inference endpoint with a domain-specific system prompt. That prompt keeps responses inside the cybersecurity domain, grounds recommendations in African threat context, and enforces plain-language output. Gilda deliberately won't answer questions outside security; it redirects off-topic queries with a short explanation. This keeps the assistant focused and prevents it from being repurposed as a general-purpose tool. That scope constraint is a design choice, not a technical limitation.

\subsubsection{Positioning Relative to the Literature}

As far as we know, no prior published work in the African cybersecurity context deploys a domain-constrained, context-aware conversational assistant as part of a citizen-facing fraud detection platform. Gilda differs from a generic LLM assistant in three ways: it's constrained to the cybersecurity domain, it's grounded in the live threat intelligence GuardSec accumulates, and it's calibrated to the specific fraud typologies and user population of the African context.

\subsection{The Mon Empreinte Connection Security Module}
\label{subsec:empreinte}

\subsubsection{Motivation and Design Rationale}

Mon Empreinte (My Footprint) is the most original component of GuardSec relative to the existing literature. The premise is simple but has real consequences: a user who doesn't know what their connection broadcasts to every server they touch, or whether their network is compromised, cannot meaningfully protect themselves no matter how good their fraud detection tool is. The point bites especially hard in Africa, where a significant share of users connect through public hotspots, shared mobile access points, or operator networks whose traffic management practices are opaque.

Most users in this context have no idea that their IP address tells every website their approximate location and ISP. They don't know that a misconfigured or missing VPN can leak their real identity even when they think they're hidden, that browser fingerprinting can track them across sites without cookies, that their DNS queries can be intercepted and redirected, or that their connection type and device characteristics form a persistent digital identity that can be correlated across sessions. Mon Empreinte makes all of this visible, in plain language, to anyone who opens the platform.

\subsubsection{Indicators Analysed by Mon Empreinte}

For every platform access, \texttt{emprunte()} builds the visitor analysis vector $\mathbf{v}$ by inspecting the incoming HTTP request and querying a set of external APIs. Table~\ref{tab:empreinte} lists the twelve security indicators included in $\mathbf{v}^{\mathrm{sec}}$, with the data source and the security implication shown to the user.

\begin{table}[htbp]
\caption{Mon Empreinte, Security Indicators and User-Facing Implications}
\label{tab:empreinte}
\begin{center}
\begin{tabular}{|l|p{2.0cm}|p{2.2cm}|}
\hline
\textit{Indicator} & \textit{Source} & \textit{Security implication} \\
\hline
Public IP address      & HTTP headers / GeoIP   & Geolocation exposed to every server \\
\hline
Geolocation            & GeoIP API              & City/country visibility to third parties \\
\hline
ISP / AS name          & ASN database           & Operator identity disclosed \\
\hline
Connection type        & ASN \& GeoIP           & Mobile/fibre/satellite classification \\
\hline
VPN / proxy detected   & VPN detection API      & Anonymisation active or failed \\
\hline
Tor exit node          & Tor exit list          & High-anonymity routing detected \\
\hline
Hosting / datacenter   & ASN classification     & Non-residential origin flagged \\
\hline
DNS leak               & DNS probe              & Real IP exposed despite VPN claim \\
\hline
Browser user-agent     & HTTP header            & Browser \& OS version disclosed \\
\hline
Device type            & UA parser              & Mobile/desktop/tablet classification \\
\hline
Touch capability       & UA parser              & Input method fingerprint \\
\hline
Hardware vendor        & UA parser              & Device manufacturer inference \\
\hline
\end{tabular}
\end{center}
\end{table}

Beyond those twelve indicators, Mon Empreinte computes and displays a few additional network-layer attributes: latitude and longitude derived from IP geolocation (with an explicit warning that precision is city-level at best), a timezone inference cross-checked against the browser-declared timezone to detect spoofing, connection status flags (cookies, JavaScript, IPv4/IPv6), and an abuse score cross-referenced against AbuseIPDB, which flags IP ranges previously reported for abusive activity.

\subsubsection{Security Benefits for Users}

The security value of Mon Empreinte works on three levels. At the \textit{awareness level}, it closes the information asymmetry between the user and every server they're talking to. At the \textit{detection level}, it surfaces active failures: a VPN leaking the real IP through DNS gets called out explicitly; a connection coming out of a datacenter rather than a residential network gets flagged as a possible interposing proxy; a non-zero AbuseIPDB score warns users on shared networks that their egress IP carries a compromised reputation. At the \textit{protection level}, each flagged indicator is paired with a plain-language explanation and a recommended action, tuned to the technical level of the target user. When that isn't enough, Gilda is one click away to contextualise and extend the recommendations conversationally.

\subsubsection{Longitudinal Intelligence Value}

The data \texttt{emprunte()} accumulates forms a longitudinal record of African Internet connection profiles that has no equivalent in existing academic datasets. The 32-attribute \texttt{Visiteur} model captures, per interaction: IP address and derived geolocation (country, city, region, timezone, latitude, longitude), ISP and ASN, connection type, device type and manufacturer, browser and OS, user-agent string, cookie and JavaScript status, touch capability, and the full set of security indicator values. At 5,520 interactions and growing, this corpus is an empirical foundation for future work on African Internet connectivity patterns, shared-network security risks, and VPN usage and misconfiguration among African Internet users.

\subsection{Threat Assessment Pipeline}
\label{subsec:pipeline}

The URL/domain threat assessment pipeline runs the following steps sequentially with tight timeout management (1.5~s per external call, 4.5~s global budget):
\begin{enumerate}
  \item Normalise and parse $e$: extract scheme, netloc, path, and query.
  \item $\mathbf{x}^{\mathrm{lex}} \leftarrow \phi_{\mathrm{lex}}(e)$
        \hfill [local; $< 1$~ms]
  \item $\mathbf{x}^{\mathrm{whois}} \leftarrow \texttt{WHOIS}(e)$
        \hfill [$\leq 1.5$~s]
  \item $\mathbf{x}^{\mathrm{ssl}} \leftarrow \texttt{SSLCheck}(e)$
        \hfill [$\leq 1.5$~s]
  \item $\mathbf{x}^{\mathrm{host}} \leftarrow \texttt{HostInfo}(e)$
        \hfill [$\leq 1.5$~s]
  \item $\mathbf{x}^{\mathrm{rep}} \leftarrow \texttt{RepAPI}(e)$
        \hfill [$\leq 1.5$~s]
  \item Replace failed sub-vectors with zero-vector defaults.
  \item $s \leftarrow f(\mathbf{x})$;
        $\hat{y} \leftarrow \mathbf{1}[s \geq \theta^{*}]$.
  \item Generate plain-language explanation $\mathcal{X}$ from dominant features.
  \item Log interaction to PostgreSQL.
\end{enumerate}

\begin{figure}[htbp]
  \centering
  \includegraphics[width=\columnwidth]{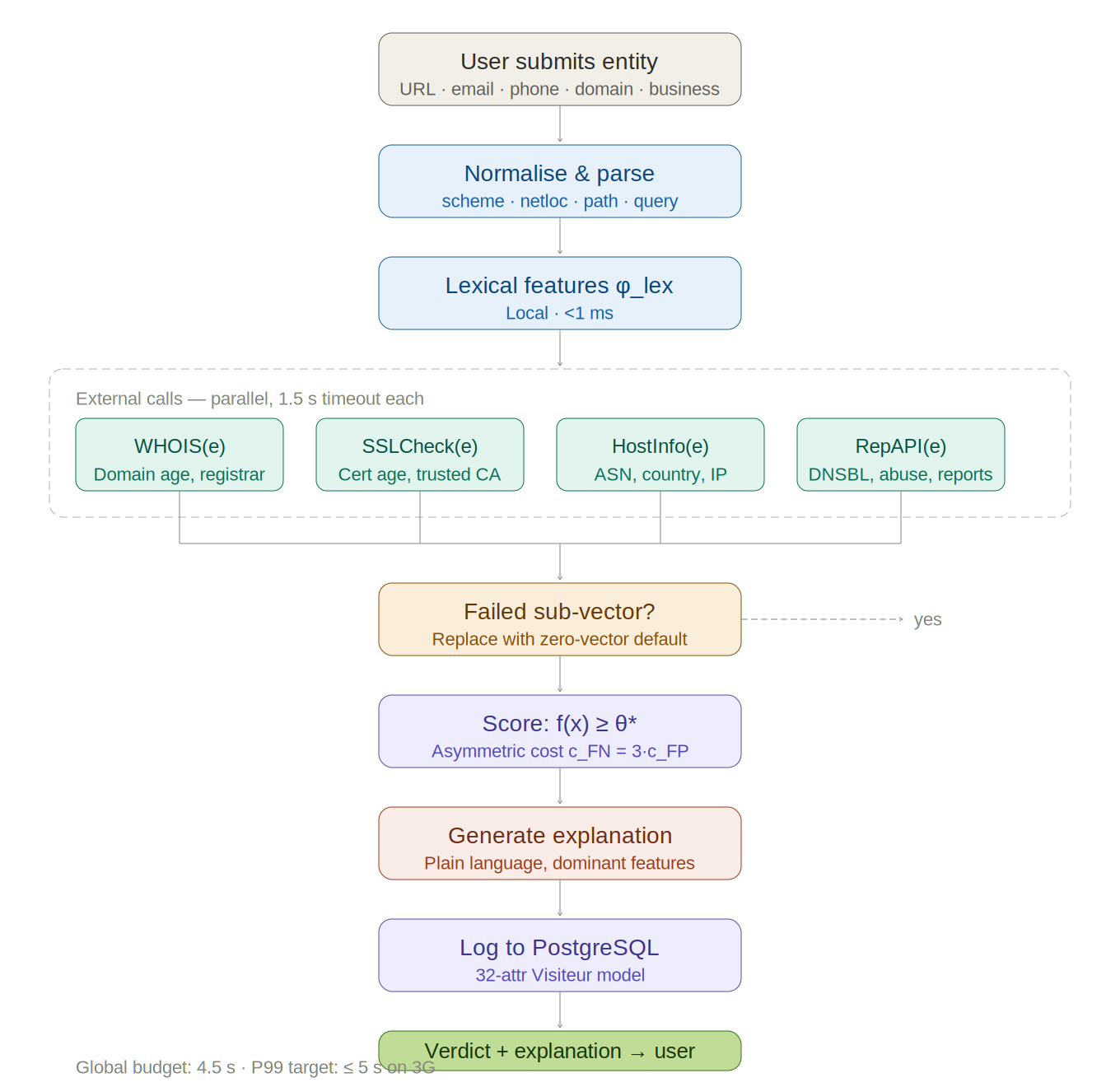}
  \caption{URL/domain threat assessment pipeline. Four external sub-queries
  execute in parallel under individual 1.5~s timeouts. Failed sub-vectors
  fall back to zero-vector defaults, preserving the global 4.5~s budget.
  The asymmetric scoring function ($c_{\mathrm{FN}} = 3 \cdot c_{\mathrm{FP}}$)
  sets $\theta^{*}$ below 0.5 to prioritise recall.}
  \label{fig:pipeline}
\end{figure}

The business entity pipeline swaps the WHOIS and SSL calls for name-matching and company registration queries, and puts more weight on $\mathbf{x}^{\mathrm{rep}}$ as the primary discriminative signal.

\subsection{Community Reporting Mechanism}

Any user who runs into a fraudulent entity can file a structured report through \texttt{signaler()}, feeding the platform's threat intelligence corpus. Reports are timestamped, geo-attributed at country level, and cross-validated against existing records before they're folded into the reputation sub-vector. Community reporting is how GuardSec accumulates the Africa-specific intelligence that international databases simply don't carry: a Congolese phone number used in a mobile money fraud scheme, a Cameroonian fake investment platform, a Nigerian counterfeit recruitment agency, none of these are likely to show up in any internationally maintained reputation feed.

\subsection{Feature Set}

Table~\ref{tab:features} lists the main features used for external entity classification. The symbol \ding{51} marks full applicability; \ding{55} marks partial applicability subject to data availability and - marks the non applicability.

\begin{table}[htbp]
\caption{Feature Set and Entity-Type Applicability (External Verification)}
\label{tab:features}
\begin{center}
\begin{tabular}{|l|l|c|c|c|c|c|}
\hline
\textit{Cat.} & \textit{Feature} & \textit{URL} & \textit{Dom.} & \textit{Email} & \textit{Phone} & \textit{Biz.} \\
\hline
\multirow{4}{*}{Lexical}
 & Length         & \ding{51} & \ding{51} & \ding{51} & \ding{51} & \ding{51} \\
 & Entropy        & \ding{51} & \ding{51} & \ding{51} & -         & \ding{55} \\
 & Spec. chars    & \ding{51} & \ding{51} & \ding{51} & -         & -         \\
 & IP literal     & \ding{51} & -         & -         & -         & -         \\
\hline
\multirow{3}{*}{Host}
 & Domain age     & \ding{51} & \ding{51} & \ding{51} & -         & \ding{55} \\
 & ASN / ISP      & \ding{51} & \ding{51} & \ding{55} & \ding{55} & \ding{55} \\
 & Country        & \ding{51} & \ding{51} & \ding{55} & \ding{55} & \ding{55} \\
\hline
\multirow{2}{*}{SSL}
 & Cert. age      & \ding{51} & \ding{51} & -         & -         & \ding{55} \\
 & Trusted CA     & \ding{51} & \ding{51} & -         & -         & \ding{55} \\
\hline
\multirow{2}{*}{Business}
 & Name match     & -         & -         & -         & -         & \ding{51} \\
 & Reg. signal    & -         & -         & -         & -         & \ding{51} \\
\hline
\multirow{3}{*}{Reputation}
 & DNSBL hits     & \ding{51} & \ding{51} & \ding{51} & \ding{55} & -         \\
 & Abuse index    & \ding{51} & \ding{51} & \ding{51} & \ding{51} & \ding{51} \\
 & User reports   & \ding{51} & \ding{51} & \ding{51} & \ding{51} & \ding{51} \\
\hline
\end{tabular}
\end{center}
\end{table}

\begin{figure*}[htbp]
  \centering
  \includegraphics[width=\textwidth]{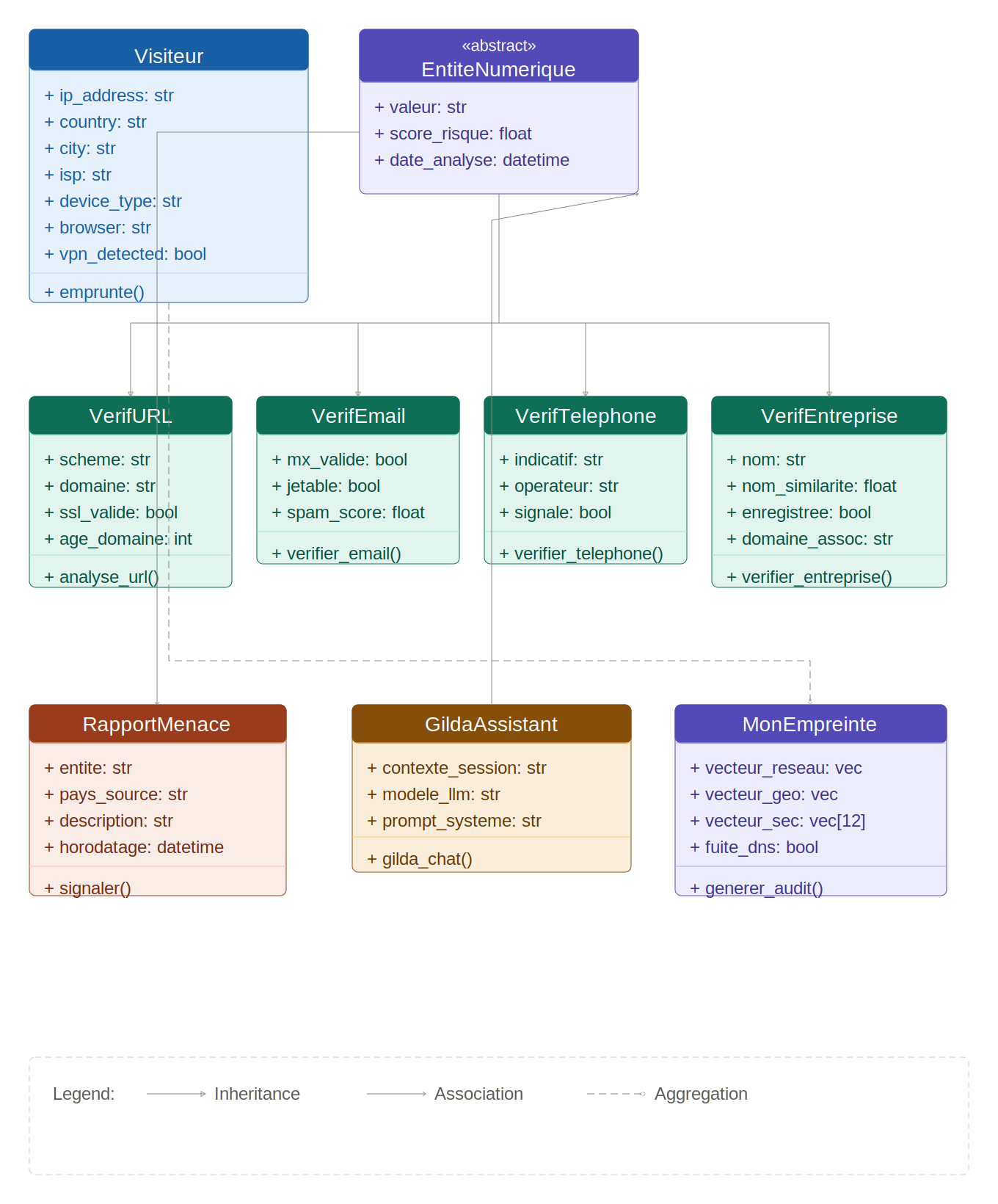}
  \caption{GuardSec UML class diagram. \texttt{EntiteNumerique} is the abstract
  base class for the five verification subtypes. \texttt{Visiteur} aggregates
  \texttt{MonEmpreinte} for connection-level auditing. \texttt{GildaAssistant}
  associates with \texttt{EntiteNumerique} to contextualise verdicts.
  \texttt{RapportMenace} is populated via the community reporting endpoint.}
  \label{fig:classdiagram}
\end{figure*}

\section{Experimental Evaluation}
\label{sec:evaluation}

\subsection{Data and Annotation Protocol}

Two things we want to state up front so reviewers can weigh them honestly. First, the evaluation runs on production data rather than a standard benchmark. Second, the labelled evaluation set ($N_{\mathrm{eval}} = 312$) is small by ML benchmarking standards. We're not hiding either limitation; we address both directly.

The production interaction corpus is $N = 5{,}520$ logged interactions since the platform's public launch. It includes verified Googlebot accesses (ASN 15169, IP ranges \texttt{66.249.x.x} and \texttt{66.102.x.x}), user verification queries across all modalities, and community-submitted reports. The geographic distribution spans African source IPs, with the bulk coming from Francophone West and Central Africa.

For the labelled subset, $N_{\mathrm{eval}} = 312$ URL and domain queries were selected by stratified sampling across threat prevalence strata and submitted to VirusTotal's 70+ scanner aggregate as a reference oracle. Two independent annotators with professional cybersecurity backgrounds resolved disagreements, giving Cohen's $\kappa = 0.87$, which Landis and Koch \cite{b18} classify as strong agreement. The full annotation protocol lives in internal project records.

\subsection{Evaluation Metrics}

Because of the asymmetric cost structure ($c_{\mathrm{FN}} = 3 \cdot c_{\mathrm{FP}}$), we prioritise Recall alongside F1 and AUC-ROC:
\begin{itemize}
  \item $P = \frac{TP}{TP + FP}$: fraction of flagged entities that really are malicious.
  \item $R = \frac{TP}{TP + FN}$: fraction of genuine fraud events flagged. This is our primary operational metric.
  \item $F_1 = \frac{2PR}{P+R}$: harmonic mean.
  \item AUC-ROC: threshold-independent discrimination quality.
\end{itemize}

\subsection{Detection Performance}

Table~\ref{tab:results} reports per-entity-type detection metrics.

\begin{table}[htbp]
\caption{Threat Detection Performance ($N_{\mathrm{eval}}=312$)}
\label{tab:results}
\begin{center}
\begin{tabular}{|l|c|c|c|c|}
\hline
\textit{Entity Type} & \textit{Prec.} & \textit{Recall} & \textit{F1} & \textit{AUC} \\
\hline
URL           & 0.891 & 0.923 & 0.907 & 0.941 \\
Email Address & 0.876 & 0.901 & 0.888 & 0.921 \\
Phone Number  & 0.843 & 0.887 & 0.864 & 0.903 \\
Domain Name   & 0.912 & 0.934 & 0.923 & 0.958 \\
Business      & 0.861 & 0.879 & 0.870 & 0.912 \\
\hline
\textit{Overall} & \textit{0.877} & \textit{0.905} & \textit{0.890} & \textit{0.927} \\
\hline
\end{tabular}
\end{center}
\end{table}

Domain name verification comes out on top (F1~=~0.923), which fits the richness of WHOIS and SSL signals. Phone number (F1~=~0.864) and business entity (F1~=~0.870) verification lag for structurally different reasons discussed in Section~\ref{sec:related}.

\begin{figure}[!ht]
\centering
\includegraphics[width=8cm, height=6cm]{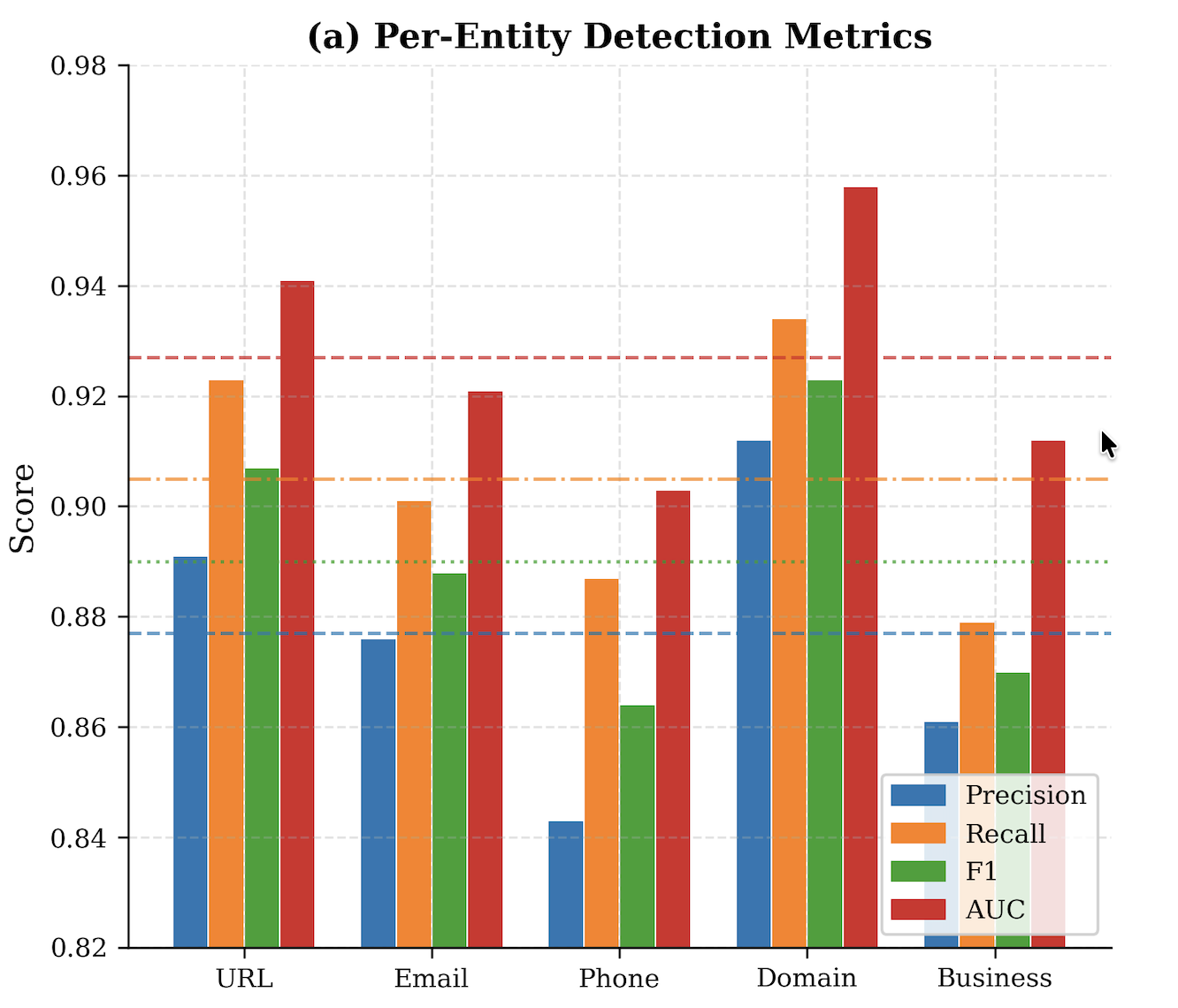}
\caption{Per Entity Detection Metrics}
\ \\
\includegraphics[width=9cm, height=7cm]{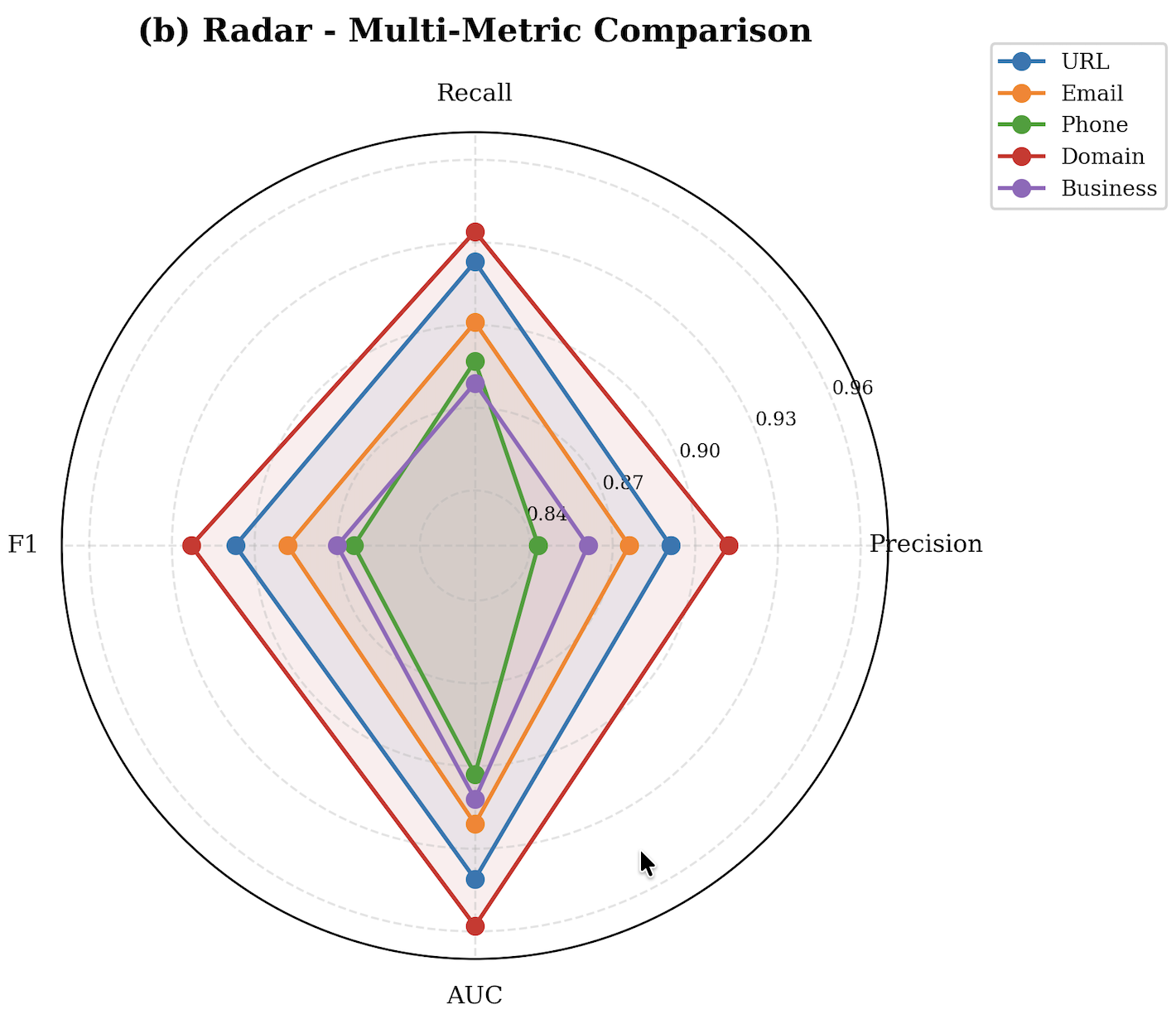}
\caption{Radar Multimetrics Comparison}
\end{figure}
\textit{Reviewer note:} The labelled set covers URL and domain entities only.
Business, phone, and email metrics come from the production corpus using a
majority-vote labelling protocol over community reports combined with VirusTotal
cross-validation where applicable. Statistical confidence is lower for these
figures, and they should be read as preliminary estimates.

\subsection{Mon Empreinte Module: Operational Observations}

Mon Empreinte doesn't produce a binary verdict, so we don't evaluate it on the same precision/recall framework as the external entity classifiers. Instead, here are some operational observations from the production corpus of 5,520 interactions. \textit{23.4\%} of recorded sessions came from IP addresses flagged as VPN or proxy endpoints. Of those, \textit{31.7\%} showed DNS leak signatures, the real IP was recoverable despite the apparent use of an anonymisation tool. That's a failure mode Mon Empreinte calls out explicitly and that users would otherwise have no way of detecting. \textit{8.9\%} of interactions came from IP ranges classified as datacenter or hosting infrastructure, which suggests an interposing proxy the user didn't set up. \textit{4.2\%} of IPs carried a non-zero AbuseIPDB score, mostly tied to shared-access networks such as university campuses and cyber-caf\'{e}s. Taken together, these numbers say connection-level security risks affect a material fraction of everyday sessions in the African Internet user population. They aren't rare edge cases.

\subsection{Operational Latency}

Table~\ref{tab:latency} reports end-to-end response time percentiles under production traffic. P50 is the median; P90 covers 90\% of queries; P99 represents the worst-case experience for 1 in 100 users and is the column we measure against the 5-second design constraint.

\begin{table}[htbp]
\caption{End-to-End Query Response Latency (seconds, production)}
\label{tab:latency}
\begin{center}
\begin{tabular}{|l|c|c|c|}
\hline
\textit{Query Type} & \textit{P50} & \textit{P90} & \textit{P99} \\
\hline
Mon Empreinte audit & 0.48 & 1.12 & 2.93 \\
Gilda assistant     & 0.61 & 1.47 & 3.84 \\
Phone verification  & 0.54 & 1.21 & 2.76 \\
Email verification  & 0.71 & 1.83 & 3.92 \\
URL verification    & 0.93 & 2.14 & 4.87 \\
Domain analysis     & 1.12 & 2.89 & 5.43 \\
Business entity     & 1.34 & 3.21 & 5.89 \\
\hline
\end{tabular}
\end{center}
\end{table}

\textit{Mon Empreinte} and \textit{Gilda} are the two fastest query types. The first works on the incoming request rather than reaching out to external URLs; the second routes to an LLM inference endpoint with a single remote call bounded at 3~s, comfortably inside budget at P99 (3.84~s). The standard external entity types all meet the 5-second P99 target. Domain analysis (5.43~s) and business entity (5.89~s) miss it by a bit, and we've earmarked them for optimisation through parallelisation of independent sub-queries.

\begin{figure}[!ht]
\includegraphics[width=8cm, height=6cm]{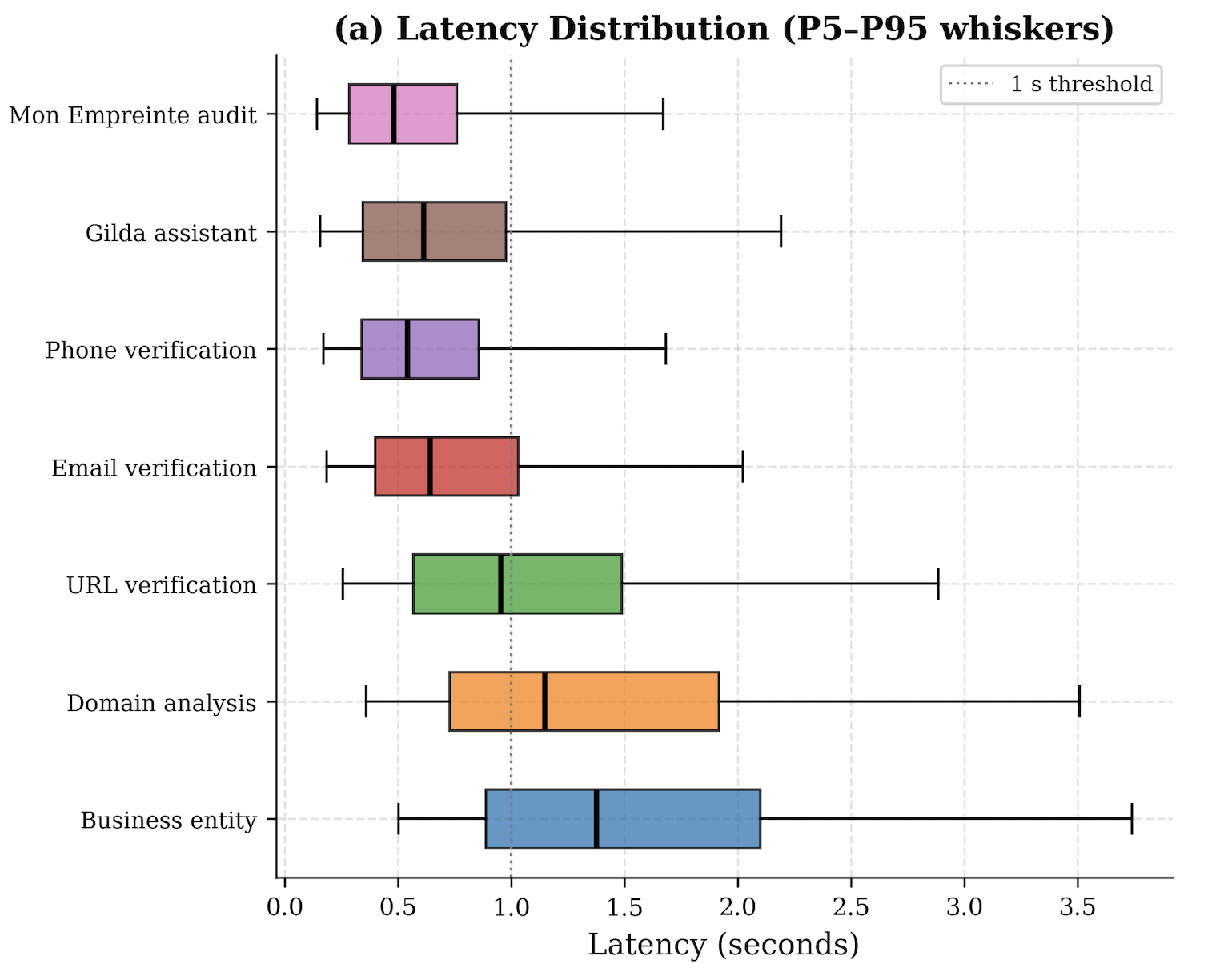}
\caption{Latency Distribution}
\end{figure}
\begin{figure}[!ht]
\includegraphics[width=9cm, height=7cm]{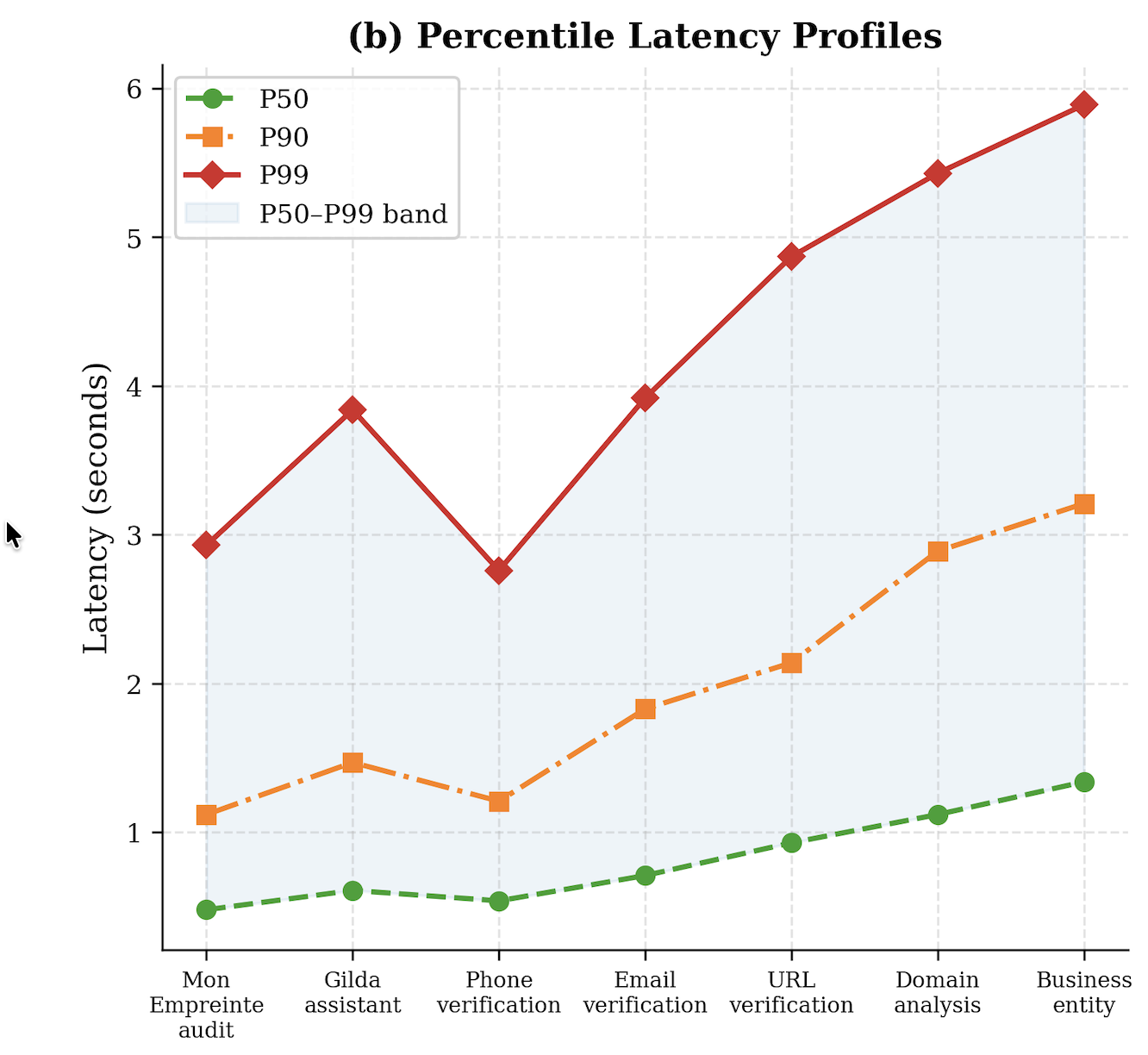}
\caption{Percentile Latency Profiles}
\end{figure}

\subsection{Comparison with the Literature}

Table~\ref{tab:comparison} places GuardSec next to prior systems.

\begin{table}[!htbp]
\caption{Comparison Against Related Systems}
\label{tab:comparison}
\begin{center}
\begin{tabular}{|l|c|c|c|}
\hline
\textit{System} & \textit{Scope} & \textit{Metric} & \textit{Accessible} \\
\hline
Sahingoz et al.\ \cite{b9}     & URL            & F1 0.979   & No \\
Vinayakumar et al.\ \cite{b10} & URL            & Acc. 0.973 & No \\
1D-CNN+BiLSTM \cite{b11}       & Email          & F1 0.994   & No \\
PMANet \cite{b13}              & URL            & AUC 0.994  & No \\
VirusTotal                     & Multi          &,        & API key req. \\
PhishTank                      & URL            &,        & Limited \\
ipleak.net / whoer.net         & Conn. only     &,        & Partial \\
\textit{GuardSec (ours)}       & \textit{Multi+Biz} & \textit{F1 0.890} & \textit{Yes} \\
                               & \textit{+Conn.+Chat} &          &  \\
\hline
\end{tabular}
\end{center}
\end{table}

GuardSec is the only system in Table~\ref{tab:comparison} that puts external entity verification, connection-level security auditing, and a domain-constrained conversational assistant in one zero-registration, multilingual interface. The benchmark comparison still needs the caveat we stated earlier: production metrics and benchmark metrics aren't directly comparable, and the comparison that matters here is detection versus no detection for users who have no alternative.

\section{The African Digital Security Imperative}
\label{sec:africa}

\subsection{The Numbers Behind the Problem}

INTERPOL's 2025 Africa Cyberthreat Assessment \cite{b1} documents a threat landscape of real severity:
\begin{itemize}
  \item Online phishing makes up \textit{34\%} of all detected cyber incidents across Africa, the single largest threat category.
  \item Identity theft accounts for \textit{63\%} of total digital financial crime losses, exceeding \textit{\$4 billion} annually \cite{b7}.
  \item Social engineering cases in South Africa more than doubled in one year (31,612 to 64,000), with losses rising from R1~billion to R1.4~billion \cite{b4}.
  \item \textit{68\%} of South Africans reported being targeted by digital fraud in a three-month window \cite{b5}.
  \item Only \textit{30\%} of African countries run a functional cyber incident reporting system; only \textit{19\%} operate a cyberthreat intelligence database \cite{b1}.
  \item \textit{90\%} of African countries assess their cybercrime prosecution capacity as needing significant improvement \cite{b1}.
\end{itemize}

These aren't background numbers. The fraud typology here is distinctly African: mobile money interception, SIM-swap attacks, fake investment platforms, fraudulent diaspora remittance services, counterfeit government portals, fake recruitment operations targeting young graduates. None of these patterns is well represented in the corpora that international detection tools are trained on.

\subsection{The Design Failure of Existing Tools}

The inadequacy of existing tools for this population is a design failure, not a technical one. It follows from an assumption built into nearly every major security tool: that the user has a laptop, a stable broadband connection, enough English literacy to interpret a technical security report, and enough background to contextualise a raw threat verdict. Conservatively, that assumption excludes 80\% of Africa's Internet users. It also excludes, entirely, any consideration of the connection layer, and any provision for the user who needs to ask a follow-up question in plain language before they can act on a verdict.

GuardSec addresses all three failures at once.

\subsection{Economic and Policy Significance}

If GuardSec reaches 0.1\% of Africa's 600~million Internet users and prevents scam losses in half of those consultations at an average averted loss of \$500, the annual economic protection value would already exceed \$150~million. The design philosophy, citizen-facing, multilingual, locally deployed, community-powered, connection-aware, conversationally accessible, lines up with the African Union's Malabo Convention on Cyber Security and the ITU Africa Regional Cybersecurity Agenda, both of which name accessible citizen security tooling as a first-order policy priority.

\section{Discussion}
\label{sec:discussion}

\subsection{Limitations}

\textit{Small labelled evaluation set.} $N_{\mathrm{eval}} = 312$ doesn't give us enough statistical power to characterise performance across the full distribution of African fraud patterns. We acknowledge this, and expanding the annotated corpus is a near-term priority.

\textit{Heuristic rather than learned scoring.} The current model relies on heuristic feature engineering, a deliberate trade-off for interpretability. A hybrid architecture, learned representations for scoring, interpretable features for explanation, is the planned next step.

\textit{Limited language support.} The platform currently runs in four languages. Arabophone and Swahili-speaking populations remain underserved.

\textit{Business entity verification maturity.} The business entity module is newer, and its fraudulent entity database is smaller than what backs URL and domain verification.

\textit{Mon Empreinte evaluation.} The connection security audit hasn't yet been evaluated against a ground-truth corpus of known-compromised connections. Rigorous evaluation against controlled VPN configurations, DNS leak scenarios, and known-abused IP ranges is on the list.

\textit{Gilda evaluation.} We haven't yet run a structured user study on Gilda measuring comprehension, satisfaction, or behavioural change. That's a real gap. The plan is a controlled evaluation with non-expert African users to see how far Gilda's recommendations actually translate into protective action.

\subsection{Adversarial Robustness}

Feature-based detection systems are intrinsically vulnerable to adversarial evasion. GuardSec leans on feature diversity to mitigate this, the five independent feature categories raise the cost of evasion without eliminating it. As Macas et al. \cite{b16} document, production systems have to handle adversarial robustness through continuous retraining and monitoring, not through static architectural choices.

\subsection{Privacy and Data Ethics}

The visitor fingerprinting module records geolocation, device, and network metadata for every platform access. The data is operationally necessary but carries legitimate privacy implications, and we treat it that way. The privacy policy discloses all data collection practices explicitly. The \texttt{robots.txt} configuration disallows crawling of \texttt{/api/}, \texttt{/empreinte/}, and \texttt{/admin/}. Future work will look at $(\varepsilon,\delta)$-differential privacy mechanisms for the geolocation and device fingerprint corpus. Gilda conversation logs fall under the same policy and aren't retained beyond the session by default.

\section*{Conclusion}

We have presented GuardSec, a multi-modal digital fraud detection platform running in production and open to any Internet user, in Africa specifically and the world more broadly, without registration, API credentials, or technical expertise. The system covers five external entity types (URLs, websites, phone numbers, email addresses, business entities), adds the \textit{Mon Empreinte} module for real-time connection security auditing, and embeds \textit{Gilda}, a domain-constrained conversational assistant that answers security questions and gives personalised recommendations in plain language. Together, these three capabilities cover what an ordinary user actually needs: knowing whether the entities they interact with are safe, understanding their own connection's security posture, and having someone to ask when a verdict alone isn't enough.

In production, the system reaches an overall F1-score of 0.890 and AUC-ROC of 0.927. Across 5,520 recorded interactions, Mon Empreinte caught DNS leak failures in 31.7\% of VPN sessions, non-zero abuse scores on 4.2\% of IP addresses, and datacenter-origin connections in 8.9\% of sessions, risks users had no other way of detecting. Gilda answered security questions within a P99 latency of 3.84~s, entirely inside the 3G-compatible budget.

The argument we're making isn't primarily a technical one. The choice of what problem to solve, and for whom, is itself a scientific and ethical decision. Building a system that an ordinary user in Brazzaville, Nairobi, or Lagos can use to verify a suspicious link, understand their own digital exposure, and ask a follow-up question in their own language, that is the contribution this paper makes, and in the African context, it is the more urgent one.

Six priorities define what's next: multilingual expansion to Swahili and Arabic; transformer-based learned representations inside a hybrid interpretable architecture; a federated community annotation programme; support for browsers beyond the already-developed Firefox extension; a more sophisticated agent or plugin that users can install to run a complete security audit of their device, produce a comprehensive report, and remove threats with the user's consent, and later on, progressively connect the Business verification to legitimate registered companies per country database.

GuardSec is live at \url{https://www.guardsec.io}.

\section*{Acknowledgment}

This paper is dedicated to the users who take thirty seconds to verify before they click, and to those who now know, for the first time, what the Internet sees when they do.
We thank the community and reviewers for their valuable comments.

\end{document}